\documentclass[epjST]{svjour}
\usepackage{graphics}
\usepackage{amssymb}
\usepackage[utf8]{inputenc}
\begin{document}
\title{Aspects of model dependence of $\eta'$-$\eta$ complex}
\subtitle{treated by going beyond the isospin limit}
\author{Davor Horvati\'c\inst{1}\fnmsep\thanks{\email{davorh@phy.hr}} \and
 Dalibor Kekez\inst{2}\fnmsep\thanks{\email{kekez@irb.hr}}
 \and Dubravko Klabu\v{c}ar\inst{1}\fnmsep\thanks{\email{klabucar@phy.hr}}}
\institute{Physics Department, Faculty of Science-PMF, University of Zagreb, Bijeni\v{c}ka cesta 32,
10000 Zagreb, Croatia \and Rugjer Bo\v{s}kovi\'c Institute, Bijeni\v{c}ka cesta 54, 10000 Zagreb, Croatia}
\abstract{Exploring the extent of model dependence, we study effects of certain
{\it Ans\"atze} for the $T$-dependence of the correction term in the QCD topological
susceptibility. The one producing  unwanted effects on results at $T > 0$
in the $\eta'$-$\eta$ complex in the usual limit of isospin symmetry, is largely 
cured from its peculiar behavior and brought into agreement with the other, when
breaking of isospin symmetry is allowed and the realistic quark mass ratio
 $m_u/m_d = 0.50$ is adopted.
} %end of abstract
\maketitle
\section{Introduction}
\label{intro}
While lattice QCD calculations now mostly agree that for the physical quark masses,
high temperatures lead to smooth, crossover restoration of chiral symmetry
around the pseudocritical transition temperature, $T\sim T_{\rm Ch}$ (for example,
see Refs. \cite{Aoki:2012yj,Buchoff:2013nra,Dick:2015twa} and references therein),
the fate of the related $U_A(1)$ symmetry restoration is still not clear
 \cite{Sharma:2018syt,Fukaya:2017wfq,Burger:2018fvb,Aoki:2012yj}.  Since the
anomalous $U_A(1)$ breaking strongly affects the  mass of the $SU(3)$ flavor-singlet
pseudoscalar meson $\eta'$, and to a lesser extent (through mixing) also the mass
of its fellow isoscalar $\eta$ from the $SU(3)$ flavor octet of light pseudoscalar
(almost-)Goldstone bosons, the temperature dependence of the $\eta'$ and $\eta$
masses are very indicative for the temperature dependence of the anomalous breaking
and of the restoration of the $U_A(1)$ symmetry of QCD.

In Ref. \cite{Horvatic:2018ztu}, we obtained a crossover $U_A(1)$ transition,
characterized by smooth, gradual melting of anomalous mass contributions.
This resulted in a significant decrease of the $\eta'$-meson mass around the
chiral transition temperature $T_{\rm Ch}$, but no decrease of the $\eta$-meson mass.
Both is consistent with the present experimental data, which indicate only the
possibility of a strong $\eta'$ mass drop  ($\gtrsim 200$ MeV) \cite{Csorgo:2009pa},
but there is no sign of any $\eta$ mass drop whatsoever.

Ref. \cite{Horvatic:2018ztu} used a phenomenologically successful effective model of
non-perturbative, low-energy QCD which, since its original inception \cite{Blaschke:2000gd},
 has been, in several variants, successfully tested in many different
phenomenological applications. This includes extending to $T > 0$
\cite{Horvatic:2007wu,Horvatic:2007qs,Horvatic:2010md,Benic:2011fv,Horvatic:2018ztu,Horvatic:2019eok}
our studies of $U_A(1)$ breaking through $\eta'$- and $\eta$-mesons at $T=0$
\cite{Klabucar:1997zi,Kekez:2000aw,Kekez:2003ri,Kekez:2005ie,Horvatic:2007mi,Benic:2014mha}.

However, regardless of how well-tested and reputable a model can be, there is 
always the issue of model and parameterization dependence of its results: is
this dependence very large, or results are largely model independent, or they
are reasonably robust at least in the qualitative sense? Exploring the extent
of model dependence was thus announced already in Ref. \cite{Horvatic:2018ztu}.
Since then, we have indeed found some examples of parameterizations which led to
$T$-dependence of $\eta'$ and $\eta$, and consequently of $U_A(1)$ (non-)restoration,
which are even qualitatively different than in Ref. \cite{Horvatic:2018ztu},
as we exemplify below. We will then also show how this is cured by adopting
realistic isospin breaking \cite{Horvatic:2020dka}.

\section{Flavorless mesons: $\eta'$, $\eta$ and % the neutral pion
 $\pi^0$  beyond isospin symmetry at $T>0$}
\label{main}

Ref. \cite{Horvatic:2018ztu} relied on our previous work \cite{Benic:2014mha} which had
demonstrated (at $T=0$, even model-independently) the soundness of the approximate way
of introducing and combining \cite{Horvatic:2007wu,Horvatic:2007qs,Horvatic:2010md,Benic:2011fv,Horvatic:2019eok,Klabucar:1997zi,Kekez:2000aw,Kekez:2003ri,Kekez:2005ie,Horvatic:2007mi}
$U_A(1)$-anomaly contributions to the masses of light quark-antiquark ($q\bar q$) flavorless
({\it i.e.,} hidden-flavor) pseudoscalar mesons, with non-anomalous,
 chiral-limit-vanishing contributions $M_{q\bar q}$ ($q=u,d,s$) to these masses.
The latter can be calculated by solving a consistently truncated system of pertinent
Dyson-Schwinger (DS) equations; first the ``gap'' equations for dressed-quark propagators
and then consistently, to ensure the correct chiral behavior, bound-state $q\bar q$
equations, typically using some effective interaction to model low-energy, nonperturbative QCD.
Other non-anomalous quantities like decay constants $f_{q\bar q}$ and condensates
 $\langle \bar q q  \rangle$, can also be calculated in such DS models.
In the present paper, our model choice is the same as in Ref. \cite{Horvatic:2018ztu}
(and as in Refs. \cite{Horvatic:2007wu,Horvatic:2007qs,Horvatic:2010md,Benic:2011fv}
preceding it and \cite{Horvatic:2019eok,Horvatic:2020dka} following it). The model
details and parameter values used in the isosymmetric limit are most conveniently
looked up in the Appendix of our Ref. \cite{Horvatic:2019eok}.

Nevertheless, the main point in the present paper will be considering the effects of the
% lightest
 quark mass model parameters breaking the isospin symmetry. Concretely,
we take from Table 3 of Ref. \cite{Horvatic:2020dka}, which performed the isospin
asymmetric refitting of our chosen model, the variants (a) $m_u = 0.67\, m_d = 4.37$ MeV,
$m_s = 115.34$ MeV, and (b) $m_u = 0.5\, m_d = 3.40$ MeV, $m_s = 115.61$ MeV.
(The variant (b) is perfectly consistent with the ratio of the QCD Lagrangian
  $u$- and $d$-quark masses: $0.493\pm 0.019$~\cite{ZylaPDG2020}.)

\begin{figure}
\resizebox{1.00\columnwidth}{!}{%
\includegraphics{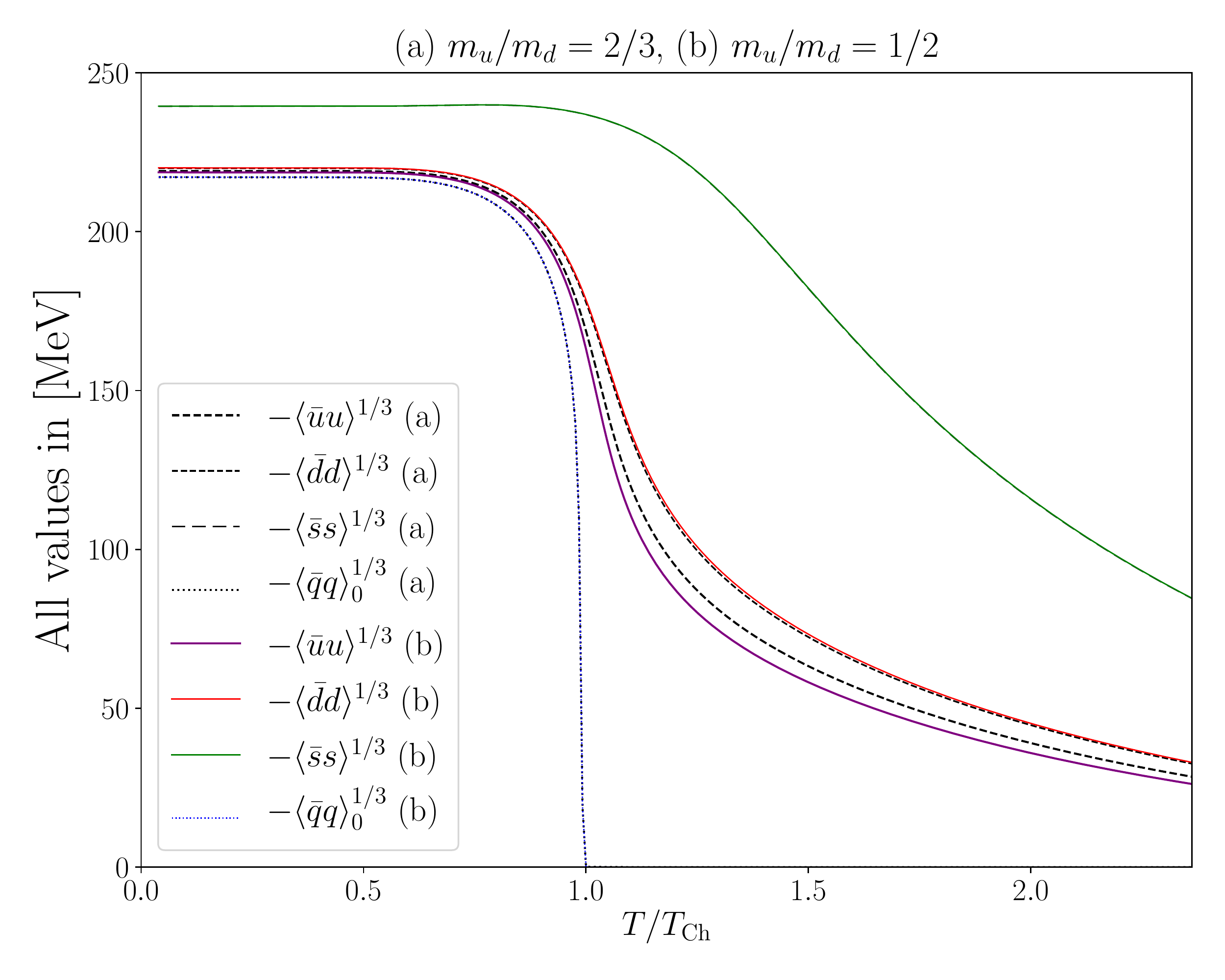} }
\caption{Temperature dependence of the condensates $\langle \bar q q  \rangle$ ($q=u,d,s$)
 for the model quark mass parameters $m_q$ in the variants \cite{Horvatic:2020dka}
 (a) and (b) -- see text. They exhibit smooth, crossover $T$-dependence, in contrast 
to the massles, chiral-limit condensate $\langle \bar q q  \rangle_0$.}
\label{fig_condensates}       % Give a unique label
\end{figure}

\begin{figure}
\resizebox{1.00\columnwidth}{!}{%
\includegraphics{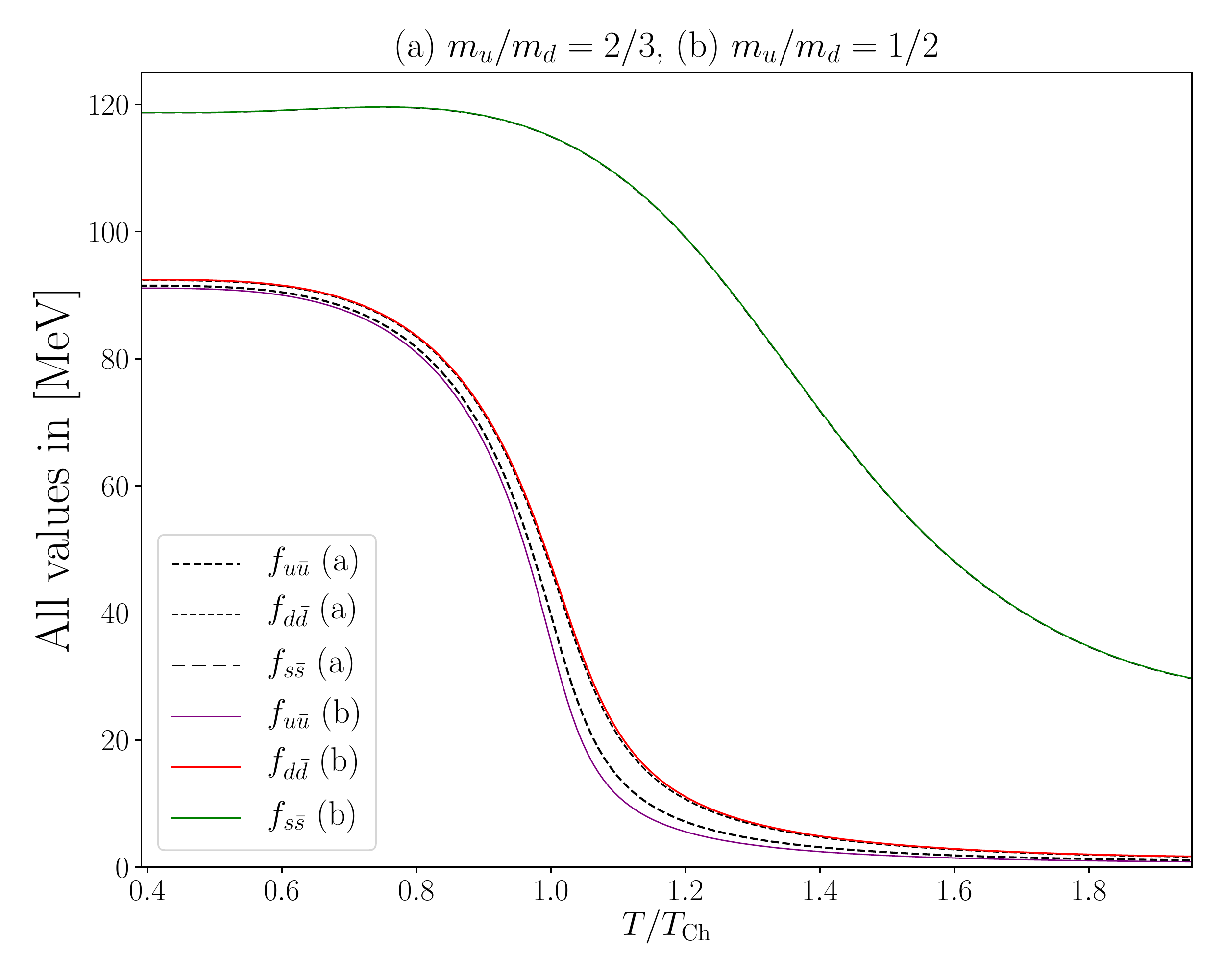} }
\caption{Temperature dependence of $q\bar q$ pseudoscalar decay constants
for $u$- and $d$-quark mass parameters breaking isospin symmetry in the
variants \cite{Horvatic:2020dka} (a) and (b), see text. }
\label{fig_decay_const}       % Give a unique label
\end{figure}

\begin{figure}
\resizebox{1.00\columnwidth}{!}{%
  \includegraphics{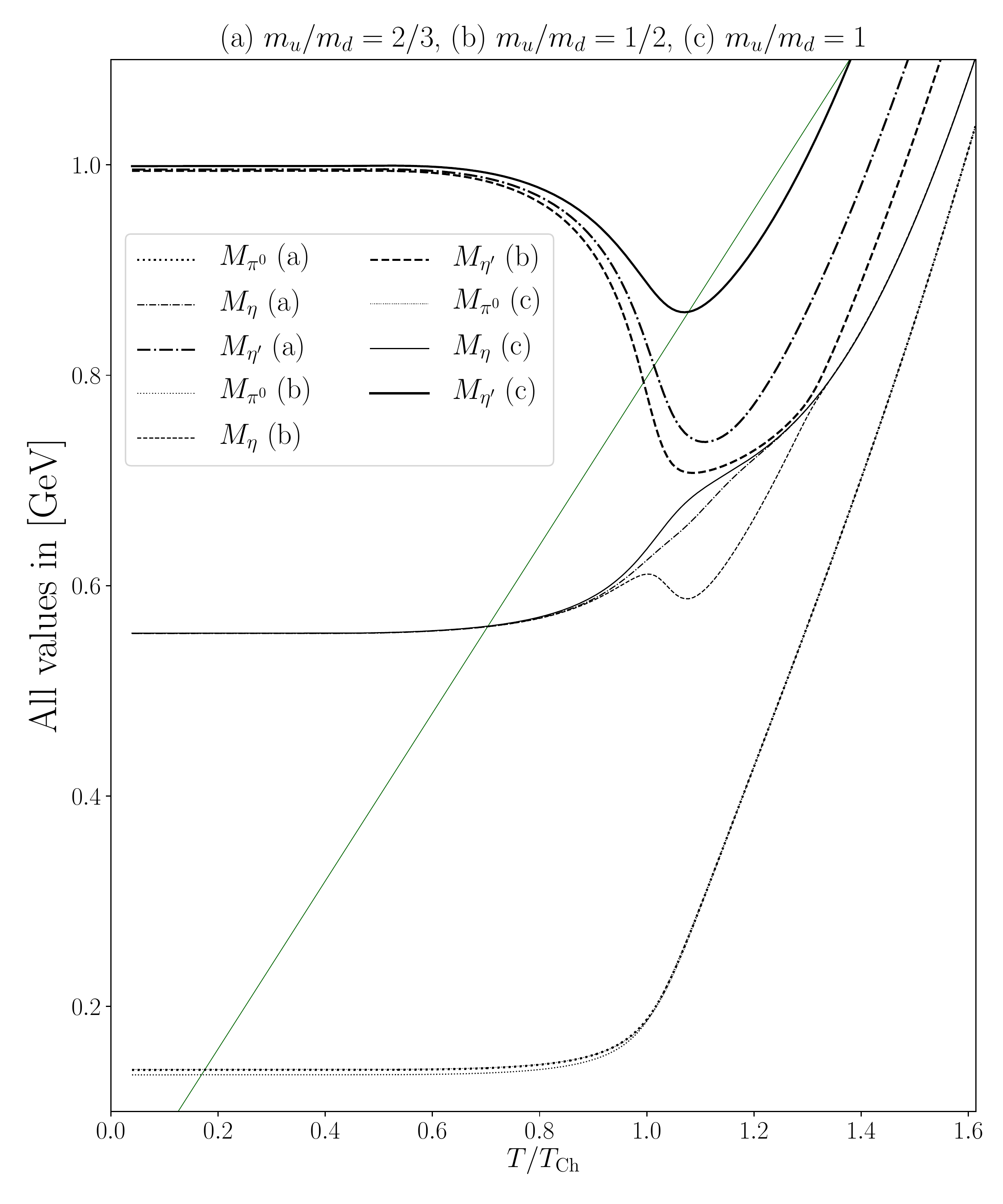} }
\caption{Temperature dependence of the masses of
% the light hidden-flavor pseudoscalar mesons
 $\eta'$, $\eta$ and $\pi^0$ for ${\cal C}_m(T)$
(\ref{casesOfC_m}) with the $\langle {\bar s}s\rangle$ condensate and $\delta = 1$.
 The isospin symmetry breaking variants  \cite{Horvatic:2020dka} (a) and (b)
are described in the text, and (c) is the usual isosymmetric variant
used also in Ref. \cite{Horvatic:2018ztu}. }
\label{fig_Cm_delta1ss}       % Give a unique label
\end{figure}

\begin{figure}
\resizebox{1.00\columnwidth}{!}{%
\includegraphics{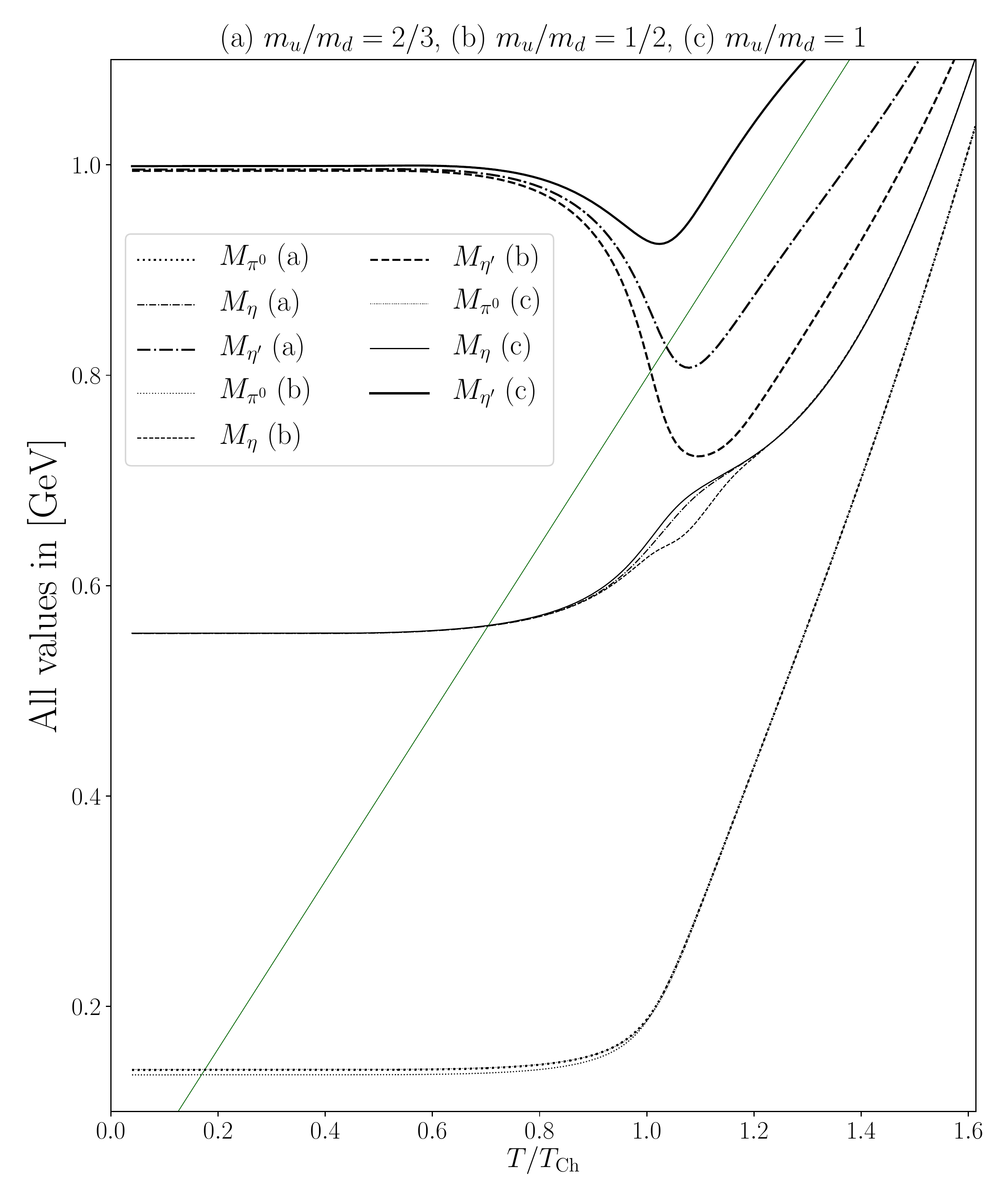} }
\caption{Temperature dependence of the masses of
% the light hidden-flavor pseudoscalar mesons
 $\eta'$, $\eta$ and $\pi^0$ for ${\cal C}_m(T)$
(\ref{casesOfC_m}) with the $\langle {\bar u}u\rangle$ condensate and $\delta = 1/5$.
 The isospin symmetry breaking variants  \cite{Horvatic:2020dka}  (a) and (b)
are described in the text, and (c) is the usual isosymmetric variant 
used also in Ref. \cite{Horvatic:2018ztu}. }
\label{fig_Cm_delta15uu}       % Give a unique label
\end{figure}

The total mass matrix (squared) ${\bf M}^2$ of the flavorless, $q\bar q$ light
pseudoscalars has the anomalous contribution ${\bf M}^2_A$. Due to the $U_A(1)$-anomaly
suppression as $1/N_c$ in the limit of the large number of colors $N_c$, ${\bf M}^2_A$
is formally treated \cite{Horvatic:2007wu,Horvatic:2007qs,Horvatic:2010md,Benic:2011fv,Horvatic:2019eok,Klabucar:1997zi,Kekez:2000aw,Kekez:2003ri,Kekez:2005ie,Horvatic:2007mi,Benic:2014mha,Horvatic:2018ztu}
as a perturbation to the non-anomalous contribution ${\bf M}^2_{N\!A}$, and,
in the first order, simply added to it: $\, {\bf M}^2 =  {\bf M}^2_{N\!A} + {\bf M}^2_{A}~. $
In the hidden-flavor basis of the unphysical pseudoscalar states $|q\bar q\rangle$ ($q=u,d,s$),
%The total mass matrix of the hidden-flavor sector in the flavor basis $|q\bar q\rangle$ is
\begin{equation}
{\bf M}^2_{N\!A}  =  \left[
\begin{array}{ccc}
 M_{{u\bar u}}^2 \, &  0 \, &  0  \\
                &     &     \\
 0 \,  &  M_{{d \bar d}}^2 \, & 0 \\
               &     &     \\
 0 \,  &  0 \,  & M_{{s\bar s}}^2 \\
\end{array}
\right] \, , \quad
 {\bf M}^2_A \, = \, \beta \, \left[
\begin{array}{ccc}
  1 \, & \,  Y \, &  \, X  \\
      &     &     \\
  Y \, & \,\, \, Y^2 \,& \,  X Y \\
      &     &     \\
  X \, &  \, \, X Y \, &  \, X^2 \\
\end{array}
\right] \, , \,\,
\mbox{where} \,\,\,\,  \beta \, = \, \frac{2 A}{f_{\pi}^2} \, , 
\label{M2NAiM2AqqbarBasis}
\end{equation}
since $\beta$ is related to the amplitudes of the $U_A(1)$ anomaly-induced
hidden-flavor sector transitions, {\it i.e.},
$|q\bar q\rangle \to |q' \bar q' \rangle $;
 see, {\it e.g.}, Figure 1 in Ref.  \cite{Horvatic:2020dka}.
%depicted schematically in Fig. \ref{Blob},
It is given by the matrix elements of the anomalous mass matrix (squared):
$\langle q\bar q | {\bf M}^2_A |q' \bar q' \rangle = b_q \, b_{q'} \, ,$
where $b_q \equiv \sqrt{\beta}$ for $q = u$. However, the amplitudes for
the transitions from, and into, heavier $s\bar s$ pairs are smaller than those
of lighter flavors. As in earlier papers \cite{Horvatic:2018ztu,Horvatic:2007wu,Horvatic:2007qs,Horvatic:2010md,Benic:2011fv,Horvatic:2019eok,Klabucar:1997zi,Kekez:2000aw,Kekez:2003ri,Kekez:2005ie,Horvatic:2007mi,Benic:2014mha},
the effects of this breaking of the $SU(3)$ flavor symmetry for $q,q' = s$
are given by $b_s = X \sqrt{\beta}$, where the flavor-breaking factor $X < 1$
is given by the ratio of the pertinent $q\bar q$ pseudoscalar decay constants,
$X = f_{u\bar u}/f_{s\bar s} < 1$. Of course, $f_{u\bar u} = f_{d\bar d} = f_\pi$ in
the limit of the $SU(2)$ isospin symmetry, which is an excellent approximation for
almost all purposes in hadronic physics. Nevertheless, we shall now consider also
its breaking, $m_d - m_u \neq 0$, however small. We allow unequal lightest quark
mass parameters and define $Y = f_{u\bar u}/f_{d\bar d}$. Then, $b_d = Y \sqrt{\beta}$
 is (at least at $T=0$) just slightly smaller than $b_u = \sqrt{\beta}$,
 just like $M_{u\bar{u}} < M_{u\bar{d}} \equiv M_{\pi^\pm} < M_{d\bar{d}}$,
even though all these differences are very small even on the scale of the pion
masses, which are the smallest of hadronic masses. But due to $Y\neq 1$, it is
no longer possible to decouple $\pi^0$ as a pure isovector without any $U_A(1)$
anomaly influence and $s\bar s$ admixture, and to isolate $\eta'$-$\eta$
 complex in their simple $2\times 2$ mass matrix; now one has to diagonalize
 the $3\times 3$ matrix 
$\, {\bf M}^2 =  {\bf M}^2_{N\!A} + {\bf M}^2_{A}\, $~(\ref{M2NAiM2AqqbarBasis})
 at every $T$.

The reason why we anyway consider the breaking of the isospin symmetry will become
clear furter below. It is related to
the way $U_A(1)$ symmetry breaking is tied to the chiral symmetry breaking in our
approach \cite{Benic:2014mha,Horvatic:2018ztu,Horvatic:2019eok,Horvatic:2020dka}
-- through the light-quark expression for the QCD topological susceptibility $\chi$
 and the corresponding full QCD topological charge parameter $A$,
\begin{equation}
\label{defA&chiShore_small_m}
A \, = \, \frac{\chi}{\, 1 \, + \, 
{\chi} \, \sum\limits_{q=u,d,s}\, \frac{1}{\, m_{q}\,\langle{\bar q}q\rangle}\, }
\,\, , \quad \mbox{with} \quad 
\chi \, = \, \frac{- \, 1}{\sum\limits_{q=u,d,s}\,\frac{1}{\,m_{q}\,\langle{\bar q}q\rangle}\,}
\, + \, {\cal C}_m \, ,
\end{equation}
where $A$ is the quantity which, in Shore's generalization \cite{Shore:2006mm,Shore:2007yn}
 of the Witten-Veneziano relation \cite{Witten:1979vv,Veneziano:1979ec}, takes place
 of the Yang-Mills topological susceptibility $\chi_{\rm{\small Y\!M}}$.
At $T=0$, $\, A = \chi_{\rm{\tiny Y\!M}} + {\cal O}({1}/{N_c}) \,$,
 so that inserting everywhere in $A$ (\ref{defA&chiShore_small_m}) the
 chiral-limit condensate $\,\langle{\bar q}q\rangle_0\,$ instead of
 the massive condensates $\,\langle{\bar q}q\rangle\,$  ($q=u,d,s$),
returns the (inverted) Leutwyler-Smilga relation \cite{Leutwyler:1992yt},
used in our original \cite{Benic:2011fv} extension of our $\eta'$-$\eta$
approach to $T>0$. But then the abrupt drop to zero of the chiral-limit
condensate $\,\langle{\bar q}q\rangle_0(T)\,$ at $T=T_{\rm Ch}$
(see Fig. \ref{fig_condensates}) causes an equally abrupt drop of not only
$\eta'$-mass, but also of $\eta$-mass, for which no indication has been found.
Contrary to that, Shore's generalization \cite{Shore:2006mm,Shore:2007yn} mandates
employing the ``massive'' condensates in Eqs. (\ref{defA&chiShore_small_m}), which
gave us, in our earlier isosymmetric $\eta'$-$\eta$ study \cite{Horvatic:2018ztu},
the smooth crossover $T$-dependence of $\eta$ and $\eta'$, with the empirically
supported \cite{Csorgo:2009pa} strong drop of the $\eta'$-mass, but no drop of
 the $\eta$-mass. 

The isosymmetric analysis \cite{Horvatic:2018ztu} was performed with the values
of model parameters which had been fixed years ago in various $T=0$ applications.
Thus, the only new parameterization was trying {\it Ans\"atze} for the unknown
$T$-dependence of the small (being of higher order in small quark masses)
correction term ${\cal C}_m$ in Eq.~(\ref{defA&chiShore_small_m}). But, its
$T=0$ value ${\cal C}_m(0)$ is fixed, as it always
 was~\cite{Benic:2011fv,Benic:2014mha,Horvatic:2018ztu,Horvatic:2020dka},
 by using Shore's \cite{Shore:2006mm} $1/N_c$-approximation 
 $A = \chi_{\mbox{\rm\scriptsize YM}}$. (As before, we adopt the lattice result
$\chi_{\mbox{\rm\scriptsize YM}} = (191\, {\rm MeV})^4$ \cite{DelDebbio:2004ns}.)
 The results in Ref. \cite{Horvatic:2018ztu} were quite insensitive to the tried
 parameterizations of ${\cal C}_m(T)$; {\it i.e.,} the resulting $T$-dependences
 of the $\eta'$ and $\eta$ obtained in Ref. \cite{Horvatic:2018ztu} were rather
 similar, although the {\it Ans\"atze} used for ${\cal C}_m(T)$ were quite different:
 the constant ${\cal C}_m(T) \equiv {\cal C}_m(0)$ $\forall \, T$ {\it vs.}
 the fit smoothly joining ${\cal C}_m(0)$ at $T\lesssim T_{\rm Ch}$ with the
 power-law $T^{-5.17}$ when $\chi(T)$ enters this regime.

Nevertheless, in our subsequent exploration of model dependences, surprising
(at first) were some of the effects of employing ${\cal C}_m(T)$ {\it Ans\"atze}
of the form (\ref{casesOfC_m}) from Ref.~\cite{Benic:2011fv},
\begin{equation}
\label{casesOfC_m}
{\cal C}_m(T) = 
{\cal C}_m(0)\left(\frac{\langle \bar{q} q\rangle(T)}{\langle \bar{q}q\rangle(0)}\right)^{\!\!\delta}
\, , \,\,
\begin{array}{c} \mbox{with the cases} \\ \,\,\,\,\,\mbox{considered\,\,\,} % \longrightarrow
 \\ \mbox{presently:} \end{array}
 \, \left\{
\begin{array}{ccc}
  q = s \, , & \delta = 1 & \mbox{in Fig.~\ref{fig_Cm_delta1ss}} \\
           &  \,&   \\
  q = u \, , & \delta = 1/5 & \mbox{in Fig.~\ref{fig_Cm_delta15uu}} \\
\end{array} \right.  \,\,\, , \quad
\end{equation}
but with the massive condensates of the {\it lightest} flavors.

Namely, if we start with the most massive and most slowly melting one,
$\langle \bar{s} s\rangle(T)$ and $\delta=1$ in Eq. (\ref{casesOfC_m}),
 the thick and thin solid curves in Fig.~\ref{fig_Cm_delta1ss}
 (the case (c)) give the respective $T$-dependences
of $\eta'$ and $\eta$ masses in the isosymmetric limit. The change with
respect to Ref.~\cite{Benic:2011fv} is qualitative and even drastic, but
it has been already explained: it just reflects going from sharp transitions
of both $\eta'$ and $\eta$ (dictated by $\langle \bar{q} q\rangle_0(T)$
dropping sharply to 0 at $T=T_{\rm Ch}$), to the smooth crossover transition
due to condensates of massive quarks. The difference with respect to
Ref.~\cite{Horvatic:2018ztu} also exists, but it is just quantitative (at least
in the most interesting region, around $T \sim T_{\rm Ch}$), {\it e.g.},
just somewhat smaller $\eta'$-mass drop, so no significant surprises here,
in Fig.~\ref{fig_Cm_delta1ss}.

Nevertheless, when Eq. (\ref{casesOfC_m}) employs the lightest and the fastest
 melting one, $\langle \bar{u} u\rangle(T)$, which is ``closest'' to
 the chiral-limit condensate $\langle \bar{q} q\rangle_0(T)$ used in
 Ref.~\cite{Benic:2011fv}, we encounter the blow-up $\eta'$-mass behavior
 unless the exponent parameter $\delta$ is an order of magnitude below 1,
 although Ref.~\cite{Benic:2011fv}, like the case with $\langle \bar{s} s\rangle(T)$
in the previous paragraph, had no problem with values as high as $\delta=1$.
 Thus we employ much smaller $\delta=1/5$ in Fig. \ref{fig_Cm_delta15uu},
 but the thick solid curve, depicting $\eta'$ mass in the isospin limit
 (again denoted as the (c)-case),
 shows significantly smaller drop around $T_{\rm Ch}$ than for the 
 ${\cal C}_m(T)$ parameterizations in Ref. ~\cite{Horvatic:2018ztu}.

 The blow-up $\eta'$-mass behavior for Eq. (\ref{casesOfC_m}) with
 $\langle \bar{u} u\rangle(T)$ is reminiscent of that encountered
 in Ref.~\cite{Horvatic:2007qs}, due to the delayed melting of the
 $U_A(1)$-anomalous contribution. The cure should thus be sought in
 faster melting of the leading term of the QCD topological susceptibility
 $\chi(T)$ which determines the anomalous mass contributions in the
 $\eta'$-$\eta$ complex. As announced in the Introduction, this can
be achieved by going realistically out of the isospin symmetry limit.
Namely, the expressions (\ref{defA&chiShore_small_m}) for $\chi(T)$,
 and consequently for $A(T)$, are dominated by the harmonic average
 of the products of the light quark masses and condensates, and a
 harmonic average is dominated by its smallest term. Here, it is the product
$m_u \, \langle \bar u u \rangle$. While the difference between the lightest
condensates $\langle \bar u u \rangle$ and $\langle \bar d d \rangle$ is barely
noticeable in Fig.~\ref{fig_condensates} at small $\, T/T_{\rm Ch}\,$, the
lighter condensate $\langle \bar u u \rangle(T)$ melts noticeably faster at
$\,T\gtrsim T_{\rm Ch}\,$. Even more important is multiplying $\langle \bar u u \rangle$
by $m_u$ now around two times smaller than $m_d$. Thus, the isospin breaking can
significantly affect the $T$-dependence of the $U_A(1)$ anomaly mass contribution
through the harmonic averages in Eqs. (\ref{defA&chiShore_small_m}).

Indeed, the dash-dotted curves and the dashed curves in both Figs.~\ref{fig_Cm_delta1ss}
and \ref{fig_Cm_delta15uu} correspond, respectively, to the intermediate
case (a) $m_u = \frac{2}{3} m_d$ and to the case (b) of the quite realistic 
ratio \cite{ZylaPDG2020} of the lightest quark mass parameters, 
$m_u = \frac{1}{2} m_d$. Thus, this improvement of the model not only
brought in a good agreement the results from two different present
 {\it Ans\"atze} (\ref{casesOfC_m}), but also in reasonable agreement
with the previous work \cite{Horvatic:2018ztu}.

\section{Summary and discussion}
\label{summary+discussion}

Although it may seem puzzling at first, it is not difficult to understand why
 the ``dominance'' (actually, exclusive presence) of the fastest-melting
condensate  $\langle \bar u u \rangle$ in the {\it Ansatz} for the correction
term ${\cal C}_m(T)$ delayed the melting of the anomalous contribution.  The
point is that ${\cal C}_m$ is always negative \cite{Benic:2011fv,Horvatic:2018ztu},
so that its particularly quick decrease left too much of the positive leading
term of $\chi(T)$, and hence of $A(T)$. This gets divided by $f_\pi(T)^2$,
which falls with $T$ faster, blowing up the $\eta'$ mass. However, this is 
avoided by introducing appropriate breaking of isospin symmetry.

Namely, we have also explained how refining \cite{Horvatic:2020dka} our model by
letting the $u$-quark mass parameter be a half of $d$-quark one, in accordance
with data \cite{ZylaPDG2020}, can compensate the effects of dominance of
$\langle \bar u u \rangle$ in the {\it Ansatz} for ${\cal C}_m(T)$ -- because
$m_u \langle \bar u u \rangle(T)$, being the smallest, will then dominate the
harmonic averages in $\chi(T)$ and $A(T)$  (\ref{defA&chiShore_small_m}),
and they decrease almost like with the previous {\it Ansatz}, and similarly
 to Ref.  \cite{Horvatic:2018ztu}.
 
The results from the two different {\it Ans\"atze} (\ref{casesOfC_m}) were brought
in a good agreement by taking $m_u = \frac{1}{2}\, m_d$ \cite{ZylaPDG2020} since in
our approach the mass of $\eta'$ has the lower limit \cite{Benic:2011fv,Horvatic:2018ztu}
at every temperature $T$, and it is $M_{s\bar s}(T)$, the non-anomalous mass of
the $s\bar s$ pseudoscalar. When $m_u = \frac{1}{2}\, m_d$, the both
 {\it Ans\"atze} (\ref{casesOfC_m}) are not very far from saturating
 this limit a little above $T_{\rm Ch}$, hence mutual similarity 
 in the case (b).

\vspace{3mm}

\section*{Acknowledgment}
D. Klabu\v{c}ar thanks for partial support to the COST Actions CA15213 THOR.

\end{document}